\documentclass[a4paper,12pt]{article}
\usepackage{eurosym}
\usepackage[dvips]{graphicx}
\usepackage{amsfonts}
\usepackage{mathrsfs}
\usepackage{graphicx} 
\usepackage{epsfig}
\usepackage{amsmath, amsthm}
\usepackage{amssymb}

\newcommand{\ba}{\begin{array}}
\newcommand{\ea}{\end{array}}

\newcommand{\beq}{\begin{equation}}
\newcommand{\eeq}{\end{equation}}
\newcommand{\ben}{\begin{enumerate}}
\newcommand{\een}{\end{enumerate}}
\newcommand{\bit}{\begin{itemize}}
\newcommand{\eit}{\end{itemize}}

\begin{document}
\title{Integrable Deformation  of Space Curves, Generalized Heisenberg Ferromagnet Equation and Two-Component Modified Camassa-Holm   Equation}
\author{Bayan Kutum\footnote{Email: bekovaguldana@gmail.com}, \, Gulgassyl  Nugmanova\footnote{Email: nugmanovagn@gmail.com},  \, \, Tolkynay Myrzakul\footnote{Email: trmyrzakul@gmail.com},  \\ Kuralay  Yesmakhanova\footnote{Email: kryesmakhanova@gmail.com} \, and Ratbay Myrzakulov\footnote{Email: rmyrzakulov@gmail.com}\\
\textsl{Eurasian International Center for Theoretical Physics and} \\ { Department of General \& Theoretical Physics}, \\ Eurasian National University,
Nur-Sultan, 010008, Kazakhstan
}
\date{}
\maketitle

\begin{abstract}
In this  paper, we provide the  geometric formulation  to the two-component Camassa-Holm equation (2-mCHE). We also study the relation between the 2-mCHE and the M-CV equation. We have shown that  these equations  arise from the invariant space 
curve flows  in three-dimensional Euclidean geometry. Using this approach we have established the    geometrical equivalence between the 2-mCHE and the M-CV equation.  The gauge  equivalence between these equations is also considered.
\end{abstract}

\section{Introduction}
In recent years there has been a growing interest in integrable  equations of the form
\begin{eqnarray}
u_{t}-u_{xxt}-f(u,u_{x},u_{xx}, u_{xxx}, ...)=0, \label{1}
\end{eqnarray}
where $u(x,t)$ is a real function and $f(u,u_{x},u_{xx}, u_{xxx}, ...)$ is some function of its arguments. This  equation admits some integrable reductions: 
the celebrated Camassa-Holm equation (CHE), the Degasperis-Procesi equation (DPE), the Novikov equation (NE) and so on. As integrable equations, all these Camassa-Holm type equations possess infinite number conservation laws, bi-Hamiltonian structures, Lax representations, peakon solutions  and other fundamental attributes of integrable nonlinear differential equations. Here we list  some  examples of the Camassa-Holm type equations.
\subsection{The CHE} The  CHE has 
the form \cite{1512.08300}
\begin{eqnarray}
q_{t}+2u_{x}q+uq&=&0,\\
q-u+u_{xx}&=&0.
\end{eqnarray}

\subsection{The modified CH equation} The modified CHE reads as \cite{1512.08300}
\begin{eqnarray}
q_{t}+[(u^{2}-u_{x}^{2})q]_{x}&=&0, \label{4} \\
q-u+u_{xx}&=&0. \label{5}
\end{eqnarray}
\subsection{The Degasperis-Procesi equation}The Degasperis-Procesi equation looks like \cite{1512.08300}
\begin{eqnarray}
q_{t}+uq_{x}+3u_{x}q&=&0,\\
q-u+u_{xx}&=&0.
\end{eqnarray}
\subsection{The Novikov  equation}The Novikov equation has the form \cite{0805.4310}
\begin{eqnarray}
q_{t}+(uq_{x}+3u_{x}q)u&=&0,\\
q-u+u_{xx}&=&0.
\end{eqnarray}
\subsection{The Hunter-Saxton   equation}The Hunter-Saxton  equation reads as  
\begin{eqnarray}
q_{t}+uq_{x}+2u_{x}q&=&0,\\
q-u_{xx}&=&0.
\end{eqnarray}
\subsection{Kaup-Boussinesq  equation}The Kaup-Boussinesq  equation is given by \cite{0906.0780}
\begin{eqnarray}
q_{t}+qq_{x}+u_{x}&=&0,\\
u_{t}-\frac{1}{4}q_{x}+\frac{1+c^{2}}{2}(uq)_{x}&=&0,
\end{eqnarray}
\subsection{Zakharov-Ito   equation}The Zakharov-Ito   equation can be rewritten as  \cite{0906.0780}
\begin{eqnarray}
q_{t}+(uq)_{x}&=&0,\\
u_{t}-4ku_{x}+u_{xxx}+3uu_{x}+qq_{x}&=&0.
\end{eqnarray}
\subsection{$b$-family equations}
There is the b-family equations of the form \cite{1510.03010}
\begin{eqnarray}
q_{t} +bqu_{x} +q_{x}u &=& 0,\\
 q +a-u+u_{xx}&=&0  
\end{eqnarray}
or 
\begin{eqnarray}
u_{t} -bau_{x} -u_{txx} + (b+1)uu_{x} - bu_{x}u_{xx} -uu_{xxx}=0, \label{18}
\end{eqnarray}
with b = 2 and b = 3, respectively.
The b-family equation can be written in the form \cite{1710.05539}
\begin{eqnarray}
(q^{1/b})_{t} + (q^{1/b}u)_{x} = 0.
\end{eqnarray}
Introducing the new function as  $p = q^{1/b}$, this implies we can define new coordinates $y, \tau$ via
\begin{eqnarray}
dy = pdy - pud\tau , \quad d\tau = dt.
\end{eqnarray}

The aim of this paper is to provide geometric formulations to the M-CV equation and to study its relation with the 2-component modified  CHE (2-mCHE). We shall show that these equations  arise from the invariant space 
curve flows  in three-dimensional Euclidean geometry.

The paper is organized as follows. In section 2, we
present  the M-CV  equation and its Lax representation (LR). In section 3, the relation between the motion of space curves and the M-CV  equation is considered. Using this relation we proved  that the Lakshmanan (geometrical) equivalent counterpart of the M-CV equation is the 2-mCHE. Some basic facts of the 2-mCHE is presented in Section 4.  The gauge equivalence between the M-CV  equation and the 2-mCHE  is considered   in Section 5. Section 6 is devoted to our concluding comments and remarks.

\section{M-CV equation}
One of examples of the peakon spin systems is the following M-CV equation
\begin{eqnarray}
 [A,A_{xt}] +(Q[A,A_{x}])_{x}+\frac{8}{\beta^{2}}A_{x}=0, \label{21}
\end{eqnarray}
where $\beta=const$ and
\begin{eqnarray}
 Q=-\frac{8\epsilon\beta^{-2}{\bf A}_{x}^{2}+{\bf A}_{x}\cdot({\bf A}\wedge{\bf A}_{tx})}{{\bf A}_{x}\cdot({\bf A}\wedge{\bf A}_{xx})}, \quad {\bf A}=(A_{1}, A_{2}, A_{3}), \label{11}
\end{eqnarray}
\begin{eqnarray}
A=\left(\begin{array} {cc} A_{3}& A^{-} \\ 
A^{+}&-A_{3}\end{array}\right), \quad A^{\pm}=A_{1}\pm iA_{2}, \quad A^{2}=I, \quad {\bf A}^{2}=1.
\end{eqnarray}
The LR of the M-CV equation  reads as
\begin{eqnarray}
\Psi_{x}&=&U_{1}\Psi,\\
 \Psi_{s}&=&V_{1}\Psi,
\end{eqnarray}
where
\begin{eqnarray}
U_{1}&=&\left(\frac{\lambda}{4\beta}-\frac{1}{4}\right)[A,A_{x}],\\
V_{1}&=&\left(\frac{1}{4\beta^{2}}-\frac{1}{4\lambda^{2}}\right)A+\left[\frac{1}{8\beta\lambda}-\frac{1}{8\beta^{2}}-\left(\frac{\lambda}{4\beta}-\frac{1}{4}\right)u\right][A,A_{x}]+\\
& + &\left(\frac{1}{2\lambda}-\frac{1}{2\beta}\right)[A,\frac{\beta}{2}A_{t}+\left(\frac{\beta u}{2}-\frac{1}{4\beta}\right)A_{x}].
\end{eqnarray}
\section{Integrable motion of space curves induced by the M-CV equation}

In this section, we  study the motion of space curves induced by the  M-CV
  equation.   Consider a smooth space curve ${\bf \gamma} (x,t): [0,X] \times [0, T] \rightarrow R^{3}$ in $R^{3}$, where $x$ is the arc length of the curve at each time $t$.  The corresponding  unit tangent vector ${\bf e}_{1}$,  principal normal vector ${\bf e}_{2}$ and binormal vector ${\bf e}_{3}$ are given by ${\bf e}_{1}={\bf \gamma}_{x}, \quad {\bf e}_{2}=\frac{{\bf \gamma}_{xx}}{|{\bf \gamma}_{xx}|}, \quad {\bf e}_{3}={\bf e}_{1}\wedge {\bf e}_{2}, $ 
respectivily. The   corresponding Frenet-Serret equation and its temporal counterpart look like 
\begin{eqnarray}
\left ( \begin{array}{ccc}
{\bf  e}_{1} \\
{\bf  e}_{2} \\
{\bf  e}_{3}
\end{array} \right)_{x} = C
\left ( \begin{array}{ccc}
{\bf  e}_{1} \\
{\bf  e}_{2} \\
{\bf  e}_{3}
\end{array} \right),\quad
\left ( \begin{array}{ccc}
{\bf  e}_{1} \\
{\bf  e}_{2} \\
{\bf  e}_{3}
\end{array} \right)_{t} = G
\left ( \begin{array}{ccc}
{\bf  e}_{1} \\
{\bf  e}_{2} \\
{\bf  e}_{3}
\end{array} \right). \label{29} 
\end{eqnarray}
Here
\begin{eqnarray}
C=
\left ( \begin{array}{ccc}
0   & \kappa_{1}     & \kappa_{2}  \\
-\kappa_{1}  & 0     & \tau  \\
-\kappa_{2}    & -\tau & 0
\end{array} \right),\quad
G=
\left ( \begin{array}{ccc}
0       & \omega_{3}  & \omega_{2} \\
-\omega_{3} & 0      & \omega_{1} \\
-\omega_{2}  & -\omega_{1} & 0
\end{array} \right),\label{30} 
\end{eqnarray}
where $\tau$,  $\kappa_{1}, \kappa_{2}$ are   torsion,  geodesic curvature and  normal curvature of the curve, respectively; $\omega_{j}$ are some real functions.  The compatibility condition of the equations (\ref{29}) reads  as
\begin{eqnarray}
C_t - G_x + [C, G] = 0\label{52} 
\end{eqnarray}
or in elements   
 \begin{eqnarray}
\kappa_{1t}- \omega_{3x} -\kappa_{2}\omega_{1}+ \tau \omega_2&=&0, \label{53} \\ 
\kappa_{2t}- \omega_{2x} +\kappa_{1}\omega_{1}- \tau \omega_3&=&0, \label{54} \\
\tau_{t}  -    \omega_{1x} - \kappa_{1}\omega_2+\kappa_{2}\omega_{3}&=&0.  \label{55} \end{eqnarray}
We now identify the spin vector ${\bf A}$ with the  tangent vector ${\bf e}_{1}$ that is we assume 
 \begin{eqnarray}
{\bf A}\equiv {\bf e}_{1}. \label{56} 
\end{eqnarray}
Let take place the following expressions    
\begin{eqnarray}
\kappa_{1}=-2\zeta, \quad \kappa_{2}=r-q, \quad \tau=-i(r+q), \label{57} 
\end{eqnarray}
where $q=-0.5(\kappa_{2}-i\tau)$ and $r=0.5(\kappa_{2}+i\tau)$ are some  functions,  $\zeta=const.$  Then we have 
\begin{eqnarray}
\omega_{1} & = &i[(0.5\lambda^{-1}-\lambda u)(q+1)+0.5\lambda^{-2}(u_{x}+u_{xx})],\label{58}\\ 
\omega_{2}&=& [(0.5\lambda^{-1}-\lambda u)(q+1)+0.5\lambda^{-1}(u_{x}+u_{xx})], \label{59} \\
\omega_{3} & = &i[0.5\lambda^{-2}-u-u_{x}].      \label{60}
\end{eqnarray}
Eqs.(\ref{53})-(\ref{55}) give us the following equations for $q, u$:
\begin{eqnarray}
q_{t}+[(u-u_{x})(v+v_{x})q]_{x}&=&0, \label{40}\\
r_{t}+[(u-u_{x})(v+v_{x})r]_{x}&=&0, \label{41}\\
q-u+u_{xx}&=&0, \label{42}\\
r-v+v_{xx}&=&0. \label{43}
\end{eqnarray}
It is the 2-mCHE.  
So, we have  proved  that the  Lakshmanan (geometrical) equivalent counterpart of  the M-CV  equation   is the 2-mCHE. 
 
 \section{Two-component modified CHE}
The two-component modified CHE (2-mCHE) has the form
\begin{eqnarray}
q_{t}+[Qq]_{x}&=&0, \label{40}\\
r_{t}+[Qr]_{x}&=&0, \label{41}\\
q-u+u_{xx}&=&0, \label{42}\\
r-v+v_{xx}&=&0, \label{43}
\end{eqnarray}
where $Q=(u-u_{x})(v+v_{x})$.
Its LR is given by \cite{1512.08300}
\begin{eqnarray}
\Phi_{x}&=&U_{2}\Phi,\\
\Phi_{t}&=&V_{2}\Phi.
\end{eqnarray}
Here\begin{eqnarray}
U_{2}&=&0.5\left ( \begin{array}{cc}
 -1  &  \lambda m  \\
-\lambda n   & 1
\end{array} \right), \\
V_{2}&=&\left ( \begin{array}{cc}
2\lambda^{-2}+0.5Q    &  -\lambda^{-1}(u-u_{x})-0.5\lambda m Q \\
\lambda^{-1}(v+v_{x})+0.5\lambda nQ    & -0.5Q 0
\end{array} \right).
\end{eqnarray}
 This LR can be rewritten as
\begin{eqnarray}
\phi_{xx}&=&(\ln{m})_{x}\phi_{x}+[0.25+(\ln{m})_{x}-0.25\lambda^{2}nm]\phi, \label{48} \\
\phi_{t}&=&(Q-\frac{u-u_{x}}{\lambda m})\phi_{x}+[2\lambda^{-2}+Q-\frac{2(u-u_{x})}{\lambda^{2}m}]\phi, \label{49}
\end{eqnarray}
where $\phi$ is  already the scalar function. Note that if $v=u$, then the 2-mCHE reduces to the mCHE (\ref{4})-(\ref{5}).

 \section{Gauge equivalence between the M-CV equation and the 2-mCHE}
Above, we have shown that the M-CV equation and the 2-mCHE are the geometrical (Lakshmanan) equivalent each to other. Lastly we note that these equations are also gauge equivalent each to other. It was shown in \cite{bayan1} that in this case $\Psi=G\Phi$.
\section{Conclusions}
 
In this work, we have studied  the relation between the M-CV equation (\ref{21}) and the 2-mCHE (\ref{40})-(\ref{43}). In particular, we have established  that the Lakshmanan (geometrical) equivalence between the  M-CV equation (\ref{21}) and the 2-mCHE (\ref{40})-(\ref{43}). Also we have noted that between these equations takes place the gauge equivalence. 
Finally we note that the above presented our results are significant for the deep understand the geometrical nature of integrable equations \cite{R13}-\cite{1907.10910}, in particular, integrable peakon systems.  
 
 \section{Acknowledgements}
This work was supported  by  the Ministry of Edication  and Science of Kazakhstan under
grants 0118ะส00935 and 0118ะส00693.

 \end{document}